\documentclass[13pt]{iopconfser}
\usepackage{setspace}

\usepackage{titlesec}
\titlespacing*{\section}{0pt}{3ex plus 1ex minus .2ex}{2ex}
\titlespacing*{\subsection}{0pt}{2.5ex plus 1ex minus .2ex}{1.5ex}
\usepackage{amsmath,amssymb,amsfonts}%
\usepackage{musicography}
\usepackage{marginnote}
\usepackage{ulem}
\usepackage{amsmath,amssymb,amsfonts}%
\usepackage{amsthm}%
\usepackage{orcidlink}
\newtheorem{theorem}{Theorem}
\newtheorem{proposition}{Proposition}%
\usepackage{mathtools}
\usepackage{mathrsfs}%
\usepackage{hyperref}
\hypersetup
    {
    colorlinks=true,
    linkcolor=blue,
    filecolor=red,      
    urlcolor=blue,
    linktoc=page,
    citecolor=magenta,
    }
    
\begin{document}

\vspace*{-3.2cm}
\title{ Trapped photon region in the phase space of sub-extremal Kerr-Newman and Kerr-Sen spacetimes}

\author{Carla Cederbaum$^{1}$ and Karim Mosani$^{2}$\orcidlink{0000-0001-5682-1033}}

\affil{$^1$Department of Mathematics, University of Tuebingen, Tuebingen, Germany}
\vspace{0.2cm}
\affil{$^2$Faculty of Mathematics, University of Vienna, Vienna, Austria}

\email{$^1$cederbaum@math.uni-tuebingen.de}
\email{$^2$karim.mosani@univie.ac.at}
\begin{abstract}
We analyse the geometry and topology of the trapped photon region in the domain of outer communication of sub-extremal Kerr-Newman and Kerr-Sen spacetimes. Specifically, we show that its projection to the (co-)tangent bundle forms a five-dimensional submanifold with topology $SO(3)\times \mathbb{R}^2$ in each setup. The proof adapts the method of Cederbaum and Jahns for sub-extremal Kerr spacetime. 
\end{abstract}
This work is a contribution to the GR24/Amaldi16 conference proceedings and will appear in the Journal of Physics: Conference Series. It presents an overview of results from an upcoming paper.
\section{Introduction}
In Schwarzschild spacetime with $M\in \mathbb{R}^+$, there exists a family of photons whose orbit under the action of the stationary Killing vector field has a constant corresponding Schwarzschild radial coordinate given by $r=3M$. Such photons are called trapped, and the set of all such trapped photons, the Trapped Photon Region (TPR) (see \S $4.1$ of \cite{Cederbaum_2019} for the definition of TPR in the domain of outer communication (DOC) of stationary spacetimes). The null geodesic associated with a trapped photon is called a trapped null geodesic. The TPR in Schwarzschild spacetime forms a timelike hypersurface, which is totally umbilic
(see \S $4.15$ of \cite{Neill_1986}) and is called a photon sphere. In the DOC of sub-extremal Kerr spacetime with $M\in \mathbb{R}^+$, i.e. $r>M+\sqrt{M^2-a^2}$ (see \S 2 of \cite{Cederbaum_2019}), trapped photons exist. 
However, unlike its static counterpart, i.e., the Schwarzschild spacetime, its TPR is not a submanifold (see Fig. 2 of \cite{Cederbaum_2019}). 

A parametrised geodesic $\gamma:A \subset\mathbb{R}\to \mathcal{M}$ in the DOC of a black hole spacetime $(\mathcal{M},g)$ (see \S 2) can be identified by an element of the tangent bundle $T\mathcal{M}$ of the spacetime in a canonical way as follows:
\begin{equation}\label{canonicalmap1}
    \mathcal{M} \supset \gamma \longmapsto \left(\gamma(0), \dot \gamma(0) \right) \in T\mathcal{M}.
\end{equation}
Here, the overhead dot denotes the derivative of $\gamma$ with respect to its affine parameter. 
Using this canonical map, one can ``lift" the set
$
P:=~\{\gamma\subset \mathcal{M} ~\vert~  \gamma: \textrm{\textit{Trapped null geodesic in the DOC}}\}
$
to the set $\Tilde{P}\subset T\mathcal{M}$. Dyatlov \cite{Dyatlov_2015} showed that in the case of Kerr spacetime, $\Tilde{P}$ is a five-dimensional submanifold of topology $SO(3)\times \mathbb{R}^2$ (see also \cite{Lili_2023}). He used the implicit function theorem on a family of slowly rotating Kerr spacetimes that he considered as perturbations of the
Schwarzschild spacetime. Cederbaum and Jahns \cite{Cederbaum_2019} proved the same result using a different, more direct approach without relying on Schwarzschild spacetime. Here, we show that one can use the latter approach to prove a similar result for two classes of stationary and axisymmetric spacetimes, whose special case is the Kerr spacetime. The first one is the Kerr-Newman spacetime
\cite{Newman_1965}, which is a solution of the Einstein-Maxwell equations. The second one is the Kerr-Sen spacetime
\cite{Sen_1992}, which is a solution of the Einstein-Maxwell-Dilaton-Axion equations. We prove the following main theorem:
\begin{theorem}\label{maintheorem}  
  In both sub-extremal Kerr--Newman and Kerr--Sen spacetimes (collectively denoted by $(\mathcal{N},h)$) that contain trapped photons, the image \(\Tilde{P} \subset T\mathcal{N}\) of the trapped photon region \(P\) (located in the domain of outer communication), under the canonical map defined in expression~(\ref{canonicalmap1}), forms a five-dimensional submanifold with topology \(SO(3) \times \mathbb{R}^2\).
\end{theorem}
This result has potential applications in several areas of mathematical relativity, including black hole uniqueness theorems \cite{Cederbaum_2016}, dynamical stability of black holes \cite{Dafermos_2008}, black hole shadows \cite{Grenzebach_2016}, and gravitational lensing phenomena \cite{Perlick_2004}. 

\section{The setup and trapped null geodesics}
Let us begin with a Lorentzian manifold $(\mathcal{M}, g)$ such that $\mathcal{M}=\Tilde{\mathcal{M}}~\backslash~\mathcal{C}_{\mathcal{M}}$, where $\Tilde{\mathcal{M}}:=\mathbb{R}\times\mathbb{R}\times S^2$, $\mathcal{C}_{\mathcal{M}}$ is the zero level set of $\rho$,  and the metric in the Boyer-Lindquist coordinates is given by ($a\in\mathbb{R}^+\cup \{0\}$)
\begin{equation*} \label{NJAmetric}
     g:=-\left(1-\frac{2f}{\rho^2}\right)dt\otimes dt+\frac{\rho^2}{\triangle}dr\otimes dr+\rho^2d\theta\otimes d\theta- \frac{4 a f S^2}{\rho^2}dt\otimes d\phi +\frac{\Sigma S^2}{\rho^2}d\phi\otimes d\phi.
\end{equation*}
Here
\begin{equation*}
   S:=\sin{\theta}, \hspace{0.2cm} C:=\cos{\theta}, \hspace{0.2cm} \rho^2 :=K+a^2 C^2, \hspace{0.2cm}
    \triangle:=K-2f+a^2, \hspace{0.2cm} \textrm{and}\hspace{0.2cm}
    \Sigma :=\left(K(r)+a^2\right)^2-a^2\triangle S^2,
\end{equation*}
where $K\geq 0$ and $f$ are free functions of $r$.
To obtain the Kerr--Newman spacetime, set $K(r)=r^2$ and $f(r)=M r-\frac{e^2}{2}$. To obtain the Kerr--Sen spacetime, set $K(r)=r^2\left(1+\frac{e^2}{Mr}\right)$ and $f(r)=M r$. By setting $e=0$ in either case, one can recover the Kerr spacetime. Here $M\in \mathbb{R}^+$ and $e\in \mathbb{R}$.  

Let $\gamma:\mathbb{R} \supseteq  I\to \mathcal{M}$ ($I$ is an open interval) be a geodesic in $(\mathcal{M},g)$. The geodesic motion in $(\mathcal{M},g)$ is completely integrable due to the presence of four first-integrals (see \S 4.1 of \cite{Neill_1995}). Two of them are associated with Killing vector fields. They are $E:=-g(\partial_t,\dot \gamma)$ and $L:=g(\partial_{\phi},\dot \gamma)$. The third first-integral is associated with the trivial Killing tensor field, i.e. the metric tensor and is given by $q:=g(\dot\gamma, \dot \gamma)$ (since we are interested in null geodesics, we have $q=0$). The fourth first-integral $\mathcal{K}$ can be shown to be associated with a Killing tensor field $\mathbb{K}$ which is analogous to the case of Kerr spacetime (see for reference \cite{Walker_1970} and also Definition 1.9 and Proposition 1.10 in \cite{Giorgi_2021}) We denote it as $\mathcal{K}:=\mathbb{K}(\dot \gamma,\dot \gamma)$. In the specific case of Kerr spacetime, $\mathcal{K}$ is called the Carter constant. Moreover, one can prove the following statement for trapped null geodesics in the DOC of $(\mathcal{M},g)$:
\begin{proposition}\label{constantradialcoordinate}
    Consider $(\mathcal{M}, g)$ admitting a black hole horizon and containing trapped photons in the DOC. If 1) $f(r=0)\leq 0$,
        2) $K=K(r)$ is a bijective function from $\mathbb{R}^+\cup \{0\}\to\mathbb{R}^+\cup \{0\}$, and 3)  $f'/K'\geq0$ (where the superscript $'$ denotes derivative with respect to $r$), then spherical photons are the only trapped photons that are contained entirely in the DOC. 
\end{proposition}
This is a generalisation of Proposition (2) of \cite{Cederbaum_2019}. In what follows, we present a unified treatment of both Kerr-Newman and Kerr-Sen spacetimes, following the approach of Cederbaum and Jahns and collectively denote the spacetimes as $(\mathcal{N},h)$.

\section{Geometry}
We now restrict to $(\mathcal{N},h)$. Due to the coordinate failure, we individually prove that $\tilde{P}\backslash\{S^2=0\}$ and a neighbourhood of $\{S^2=0\}$  are five-dimensional submanifolds of $T\mathcal{N}$.
Let us call the coordinates $(t,r,\theta,\phi, \dot t, \dot r,\dot \theta,\dot \phi)$ of $T\mathcal{N}$ as simply $(x^i)$, where $1\leq i\leq 8$. The relation between these coordinates for a trapped null geodesic $\gamma\subset \mathcal{N}$ (with associated first-integrals $E$, $L$, and $\mathcal{K}$) is described by the following set of first-order geodesic equations (where $n\in\{2,3,5\}$):
\begin{align*}
x^6 &=0,\\
\left(x^7\right)^2&=f_{7}\left(x^n, L, \mathcal{K}\right),\\
x^8 &=f_{8}\left(x^n, L, \mathcal{K}\right).
\end{align*}
Three points should be highlighted here:
1) The first equation above
follows from Proposition (\ref{constantradialcoordinate}); 
2) $(\mathcal{N},h)$ is stationary and axisymmetric, and hence the geodesic motion does not depend on the coordinates $x^1$ and $x^4$;
3) The geodesic motion does depend on the first-integral $E$ but the set of geodesic equations is invariant under the scaling $(E,x^j)\to \lambda (E,x^j)$, where $5\leq j\leq 8$, and hence $E$ can be replaced by $x^5$ (and other terms that are absorbed in the functions). This replacement also reduces the number of equations from four to three.

We now construct a map
$
\mathcal{F}: T\mathcal{N}\backslash \{S^2=0\} \to \mathbb{R}^3,
$
of the form
$
\mathcal{F}(z)=\left(F_{1}(z),F_{2}(z),F_{3}(z)\right),  ~z\in T\mathcal{N}\backslash \{S^2=0\}
$
where
$F_k: T\mathcal{N}\backslash \{S^2=0\}\to \mathbb{R}
$ 
($1\leq k\leq 3$) are smooth functions  defined as (where $n\in\{2,3,5\}$)
\begin{align*}
 F_1 & =  x^6,\\ 
  F_2   &=\left(x^7\right)^2-f_{7}\left(x^n,L,\mathcal{K}\right),\\
  F_3 & =  x^8 -f_{8}\left(x^n,L, \mathcal{K}\right).\\
\end{align*}
 Such a construction ensures that $\mathcal{F}$ maps $z\in T\mathcal{N}\backslash \{S^2=0\}$ to $0\in\mathbb{R}^3$ if and only if $z\in \tilde{P}\backslash\{S^2=0\}\subset T\mathcal{N}\backslash \{S^2=0\}$. We then prove that the differential map
\begin{equation*}
    d_z\mathcal{F}: T_z \left(T\mathcal{N}\backslash \{S^2=0\}\right) \xrightarrow{\sim} T_{\mathcal{F}(z)}\mathbb{R}^3,~z\in T\mathcal{N}\backslash \{S^2=0\}
\end{equation*}
 is surjective $\forall$ $z\in \tilde{P}\backslash\{S^2=0\}$. 
 Finally, we use the submersion theorem to prove that  
$
\mathcal{F}^{-1}\left(0\in \mathbb{R}^3\right)\subset T\mathcal{N}
$
is a submanifold of dimension $8-3=5$. 

To prove similar result for a neighbourhood of $\{S^2=0\}$ in $\tilde{P}$, we first 
construct a map
$
\mathcal{H}: T\mathcal{N} \to \mathbb{R}^3
$
of the form
$
\mathcal{H}(z)=(H_{1}(z) , H_{2}(z), H_{3}(z)),
$  
$z\in T\mathcal{N},
$
where
$H_k: T\mathcal{N}\to \mathbb{R}
$ 
($1\leq k\leq 3$) are defined as
\begin{align*}
    H_1 &=q, \\
    H_2 &= \mathcal{K}-E^2 ~Q_{\textrm{trap}}(r),\\
    H_3 &= L-E ~\Phi_{\textrm{trap}}(r).
\end{align*}
Here $Q_{trap}(r):=\mathcal{K}/E^2$ and $\Phi_{trap}(r):=L/E$ are the normalised first-integrals corresponding to a trapped null geodesic at radial coordinate $r$ (we can show that for trapped photons in the DOC of $(\mathcal{N},h)$, $E\neq 0$). $\mathcal{H}$ as defined above ensures that $\mathcal{H}^{-1}(0\in\mathbb{R}^3)\subset T\mathcal{N}$ is the canonical projection of trapped photon region $P$. We then prove that $d_z\mathcal{H}$ is surjective $\forall$ $z\in \tilde{P}\cap\{S^2=0\}\subset T\mathcal{N}$. This ensures (using the submersion theorem) that there exists a neighbourhood of $\Tilde{P}\cap \{S^2=0\}$ in $\Tilde{P}$  which is a submanifold of $T\mathcal{N}$ of dimension $8-3=5$. Summing up, we conclude that $\tilde{P}\subset T\mathcal{N}$ is a five-dimensional submanifold.

\section{Topology}
We now prove that the topology of $\tilde{P}^{\musFlat{}}\subset T^*\mathcal{N}$ is $SO(3)\times \mathbb{R}^2$. Here $\tilde{P}^{\musFlat{}}\subset T^*\mathcal{N}$ is the image of $\tilde{P}\subset T\mathcal{N}$ under the musical isomorphism. We consider the first-integral equations as before, but now in terms of the coordinates $(x_i)$, where $1 \leq i\leq 8$, of the cotangent bundle. It turns out that $\tilde{P}^{\musFlat{}}$ is invariant under time translation and rescaling of $E$, i.e. $(E,x_{j})\mapsto \lambda (E,x_{j})$ where $5\leq j \leq 8$. One of the first-integral equations is $x_{5}=-E$. We therefore restrict our attention to a six-dimensional slice 
$
\{x_{1}=0, x_{5}=-1\}.
$
of $T^*\mathcal{N}$ and prove that the intersection of the trapped photon region with this slice 
$$
X:=\tilde{P}^{\musFlat{}}\cap\{x_{1}=0, x_{5}=-1\},
$$ 
has topology $SO(3)$. The main steps involved are as follows: First, we divide $X$ into three components as $X=U_E~\cup U_N~\cup U_S$, where 
\newline
$$U_E:=X\backslash \{S^2=0\}, ~U_N:=X\cap\{S^2<\epsilon;~\theta<\pi/2\}, ~\textrm{and} ~U_S:=X\cap\{S^2<\epsilon;~\theta>\pi/2\},$$
\newline
and $\epsilon>0$ is a suitable positive number. Then we prove that $$U_E\approx (0,\pi)\times \mathbb{S}^1\times \mathbb{S}^1, ~U_N\approx \mathbb{S}^2\cap \{S^2<\epsilon;~\theta<\pi/2\}\times \mathbb{S}^1, \textrm{and} ~U_S\approx \mathbb{S}^2\cap \{S^2<\epsilon;~\theta>\pi/2\}\times \mathbb{S}^1.$$

Next, we employ the Seifert–van Kampen theorem first on $U_N$ and $U_E$, and then on $U_N \cup U_E$ and $U_S$ to finally obtain $\pi_1(X)=\mathbb{Z}_2$ (where $\pi_1$ stands for the first fundamental group).  We then use the classification results of closed three-manifolds (see e.g. \cite{Hatcher_2018}) to conclude that $X\approx L(2,1) \approx SO(3)$. Finally, $\tilde{P}^{\musFlat{}}\approx X\times \mathbb{R}^2 \approx SO(3)\times \mathbb{R}^2$. 

\section{Technical Distinctions from the Sub-Extremal Kerr Case}
The generalisation from sub-extremal Kerr to sub-extremal 
Kerr–Newman and Kerr–Sen spacetimes follows the same overall strategy but requires addressing several key distinctions. These primarily involve verifying the surjectivity of the differential maps
$d_z\mathcal{F}$ and $d_z\mathcal{H}$, and analysing the topological structure of the regions $U_E$, $U_N$, and $U_S$. These differences arise due to differences in the functional forms of $f(r)$ and $K(r)$.

A crucial ingredient is examining the behaviour of the coordinate $x^3$ along null geodesics in the domain of outer communication (DOC). This requires adapting results from \S 4.5 of O'Neill \cite{Neill_1995} to the Kerr–Newman and Kerr–Sen settings. We can show that all the necessary results from this section can be proven to be true for $(\mathcal{M},g)$.

Next, we need to prove that the first-integral $\mathcal{K}$ is associated with a Killing tensor field whose expression is known explicitly (similar to the result by Penrose and Walker in the Kerr spacetime case \cite{Walker_1970}). We can show that the statement holds true for $(\mathcal{M},g)$ in general. This requirement is essential to carry out the proof of the second part of \S 3 here.

Further, we need the extension of key results from  \S 3 and the early part of \S 4 of Teo's analysis \cite{Teo_2003}, including a variant of Figure 2 in their paper to $(\mathcal{N},h)$. In addition to the above, we establish the non-existence of trapped photons with zero energy $E=0$ in the DOC of $(\mathcal{N},h)$. This result justifies the normalisation of the first-integrals $\mathcal{K}$ and $L$ for any trapped null geodesic, which we denote respectively as $Q_{trap}$ and $\Phi_{trap}$. Both these analyses are essential for explicitly computing entries of the $8\times 3$ matrices representing the linear maps $d_z\mathcal{F}$ and $d_z\mathcal{H}$. These computations give us a contradiction when assuming that the rank of the matrix is less than three, thus completing the argument for surjectivity. Finally, the generalisation of Teo’s results and \S4.5 of O'Neill to $(\mathcal{N},h)$ is also used to determine the topological structure of $U_N$, $U_S$, and $U_E$.

\section*{Acknowledgements}
This research was funded in part by the Austrian Science Fund (FWF) [grants DOI \href{https://www.fwf.ac.at/en/research-radar/10.55776/EFP6}{10.55776/EFP6}]. For open access purposes, the authors have applied a CC BY public copyright license to any author-accepted manuscript version arising from this submission.
\bibliography{iopart-num}

\end{document}